\documentclass[twocolumn]{article}

\usepackage[english]{babel}
\usepackage[utf8x]{inputenc}
\usepackage[T1]{fontenc}
\usepackage{graphicx}
\usepackage{listings}
\usepackage[top=2cm, bottom=2cm, left=1.5cm , right=1.5cm]{geometry}
\usepackage{amsmath}
\usepackage{amsfonts}
\usepackage{hyperref}
\usepackage[numbers,compress]{natbib}

\newcommand{\tss}[1]{\textsuperscript{#1}}
\def\wr{\omega_r}
\def\w{\omega}
\def\g{\gamma}
\def\l{\lambda}

\renewcommand{\Re}{\mathrm{Re}}
\renewcommand{\Im}{\mathrm{Im}}

\begin{document}
\title{Oscillation threshold of a clarinet model : a numerical continuation approach}

\author{Sami Karkar\footnote{karkar@lma.cnrs-mrs.fr}
\and Christophe Vergez
\and Bruno Cochelin\\
Laboratory of Mechanics and Acoustics, CNRS -- UPR 7051\\
31 chemin J. Aiguier, 13402 Marseille Cedex 20, France\\
Aix-Marseille University, 3 place Victor Hugo - 13331 Marseille Cedex 03, France\\
\'Ecole Centrale Marseille\\ P\^ole de l'\'Etoile, Technop\^ole de Ch\^ateau-Gombert\\ 38 rue Frédéric Joliot-Curie, 13451 Marseille Cedex 20, France}

\date{December, 20th, 2010}

\maketitle

\begin{abstract}
This paper focuses on the oscillation threshold of single reed instruments. Several characteristics such as blowing pressure at threshold, regime selection, and playing frequency are known to change radically when taking into account the reed dynamics and the flow induced by the reed motion. Previous works have shown interesting tendencies, using analytical expressions with simplified models. In the present study, a more elaborated physical model is considered. The influence of several parameters, depending on the reed properties, the design of the instrument or the control operated by the player, are studied. Previous results on the influence of the reed resonance frequency are confirmed. New results concerning the simultaneous influence of two model parameters on oscillation threshold, regime selection and playing frequency are presented and discussed. The authors use a numerical continuation approach. Numerical continuation consists in following a given solution of a set of equations when a parameter varies. Considering the instrument as a dynamical system, the oscillation threshold problem is formulated as a path following of Hopf bifurcations, generalising the usual approach of the characteristic equation, as used in previous works. The proposed numerical approach proves to be useful for the study of musical instruments. It is complementary to analytical analysis and direct time-domain or frequency-domain simulations since it allows to derive information that is hardly reachable through simulation, without the approximations needed for analytical approach.
\end{abstract}



\section{Introduction}
Woodwind instruments have been intensively studied since Helmholtz in the late 19\tss{th} century. In particular, the question of the oscillation threshold has been the focus of many theoretical, analytical, numerical and experimental studies\cite{backus:1963,benade:1968,saneyoshi:1987,fletcher:livre,fletcher:1993,chang:1994,tarnopolsky:2000}. Following the works of Backus and Benade, Worman\cite{worman:these} founded the basis of the small oscillation theory for clarinet-like systems, and Wilson and Beavers\cite{wilson:1974} (W\&B) greatly extended his work on the oscillation threshold.

Thompson\cite{thompson:1979} mentioned the existence of an additional flow, related to the reed motion, and the importance of the reed resonance on intonation as early as 1979, but only recent literature such as Kergomard and Chaigne\cite{kergomard:livre} (chap. 9), Silva et al.\cite{silva:2008} and Silva\cite{silva:2009} extended the previous works of Wilson and Beavers to more complex models, taking into account the reed dynamics and the reed-induced flow.

Whereas the static regime of the clarinet (i.e. when the player blows the instrument but no note is played) can be derived analytically, the stability of this regime and the blowing pressure threshold at which a stable periodic solution occurs is not always possible to write analytically, depending on the complexity of the model under consideration. In the latter two references, authors analyse the linear stability of the static regime through a characteristic equation. It is a perturbation method based on the equations written in the frequency domain, that suppose an arbitrarily small oscillating perturbation (with unknown angular frequency $\w$) around the static solution. Substituting the linear expressions of the passive components -- involving the acoustical impedance $Z$ defined as: $P(\w)$=$Z(\w) U(\w)$ and the reed's mechanical transfer function $D$ defined as: $X(\w)$=$D(\w) P(\w)$, where $P(\w),U(\w)$ are the acoustical pressure and volume flow at the input end of the resonator and $X(\w)$ is the reed tip displacement -- into the coupling equation linearised around the static solution ($U = \left.f^{lin}\right|_{static}(X,P)$) and balancing the first harmonic of the oscillating perturbation, one gets the characteristic equation that the angular frequency $\w$ and the blowing pressure $\g$ must satisfy at the instability threshold.

As pointed out by Silva\cite{silva:2009}, the convergence towards an oscillating regime with angular frequency $\w$ after destabilisation of the static regime is not certain and requires to consider the whole non linear system and compute periodic solutions beyond the threshold, in order to determine the nature of the Hopf bifurcation (direct or inverse). However, in the case of a direct bifurcation, the oscillation threshold (for a slowly increasing blowing pressure) corresponds to the solution of the characteristic equation with the lowest blowing pressure $\g_{th}$. Periodic oscillations emerge at this point with an angular frequency $\w_{th}$ that depends only on the other model parameters values.

In the present paper, we propose a different method for computing the oscillation threshold that proves to be faster, more general and more robust: numerical continuation. Numerical continuation consists in following a given solution of a set of equations when a parameter varies. To the authors knowledge, it is the first time this numerical method is applied to the physics of musical instruments. While time-domain or frequency-domain simulations allow to study complex physical models with at most one varying parameter, the method proposed in this work allows to investigate the simultaneous influence of several parameters on the oscillation threshold.

The general principle of the method is the following : we first follow the static regime while increasing the blowing pressure, all other parameters being kept constant, and detect the Hopf bifurcations. Then, we follow each Hopf bifurcation when a second parameter $\mu$ is allowed to vary ($\mu$ being, for instance, the reed opening parameter or the reed resonance frequency). Finally, the resulting branch of each bifurcation is plotted in the plane $(\mu,\g_{th})$ for determination of the oscillation threshold and of the selected regime, and in the plane $(\mu,\w_{th})$ to determine the corresponding playing frequency.

The paper is structured as follows. In section \ref{sec:model}, a physical model of a clarinet is reviewed. The method is described in section \ref{sec:cont}, defining the continuation of static solutions and Hopf bifurcations. In section \ref{sec:results}, the results are reviewed. First, basic results on the reed-bore interaction are compared with the previous works (W\&B\cite{wilson:1974} and Silva et al.\cite{silva:2008}). Then, extended results are presented, showing how the control of the player and the design of the maker influence the ease of play and the intonation.

\section{Physical model of single reed instruments\label{sec:model}}
The model used in this study is similar to the one used in Silva et al.\cite{silva:2008}. It is a three-equation physical model that embeds the dynamical behaviour of the single reed, the linear acoustics of the resonator, and the nonlinear coupling due to the air jet in the reed channel. It assumes that the mouth of the player (together with the vocal tract and lungs) provides an ideal pressure source, thus ignoring acoustic behaviour upstream from the reed.

\subsection{Dynamics of the reed}
The reed is indeed a three-dimensional object. However, due to its very shallow shape, it can be reduced to a 2D-plate or even, considering the constant width, a 1D-beam with varying thickness. Detailed studies of such models have been conducted (see for instance Avanzini and van Walstijn\cite{avanzini:2004} for a distributed 1D-beam model) and lumped models have been proposed. Mainly, two lumped models are widely used:
\begin{itemize}
\item a one degree of freedom mass-spring-damper system, accounting for the first bending mode of the reed,
\item a simpler one degree of freedom spring, assuming a quasi-static behaviour of the reed.
\end{itemize}

Most papers studying clarinet-like systems use the second model. This kind of simplification is made assuming that the reed first modal frequency is far beyond the playing frequency, thus acting as a low-pass filter excited in the low frequency range, almost in phase. However, some authors (see for instance W\&B\cite{wilson:1974} or Silva\cite{silva:2009}) showed that the reed dynamics, even rendered with a simple mass-spring-damper oscillator, cannot be ignored because of its incidence on the physical behaviour of the dynamical system.

Thus, in the present study, the reed dynamics is rendered through a lumped, one-mode model. It reflects the first bending mode of a beam-like model with a reed modal angular frequency $\wr$, a modal damping $q_r$, and an effective stiffness per unit area $K_a$.

The reed is driven by the alternating pressure difference $\Delta P$=$P_m-P(t)$ across the reed, where $P_m$ is the pressure inside the mouth of the player and $P(t)$ is the pressure in the mouthpiece. Given the rest position of the tip of the reed $h_0$ and the limit $h=0$ when the reed channel is closed, the tip of the reed position $h(t)$ satisfies the following equation:
\begin{equation}\label{eq:reed}
\frac{1}{\wr^2}h''(t) = h_0-h(t) - \frac{q_r}{\wr}h'(t) - \frac{\Delta P}{K_a}
\end{equation}

Contact with the mouthpiece could also be modelled, using a regularised contact force function (as presented by the authors in a conference paper\cite{karkar:ica2010}). It will not be discussed or used in this paper, as large amplitude oscillations will not be discussed. However, it is important to note that such phenomenon cannot be ignored when the whole dynamic range of the system is under study.

\subsection{Acoustics of the resonator}
The acoustic behaviour of the resonator of a clarinet is here assumed to be linear. It is modelled through the input impedance $Z_{in}$=$P/U$, where $P$ is the acoustical pressure and $U$ the acoustical volume flow at the reed end of the resonator. To keep a general formulation, we use a (truncated) complex modal decomposition of the input impedance (as described by Silva\cite{silva:2009}) of the form:
\begin{equation}\label{eq:imped_freq}
Z_{in}(\w) = \frac{P(\w)}{U(\w)} = Z_c \sum_{n=1}^{N_m} \left( \frac{C_n}{j\w-s_n} + \frac{C^*_n}{j\w-s^*_n} \right).
\end{equation}
This leads, in the time domain, to a system of $N_m$ differential equations, each governing a complex mode of pressure at the input of the resonator :
\begin{equation}\label{eq:imped}
P_n'(t) = C_n Z_c U(t) + s_n P_n(t),
\end{equation}
where $Z_c=\rho c/S$ is the characteristic impedance of the cylindrical resonator, whose cross-sectional area is $S$.
The total acoustic pressure is then given by :
\begin{equation}\label{eq:ptot}
P(t) = 2\sum\limits_{n=1}^{N_m} \Re\big(P_n(t)\big).
\end{equation}

In practice, the modal coefficients $(C_n,s_n)$ can be computed in order to fit any analytical or measured impedance spectrum to the desired degree of accuracy (depending on the total number of modes $N_m$).

In the present study, we use an analytical formulation for a cylinder of length $L=57cm$ (if not stated otherwise), radius $r=7mm$, taking into account the radiation at the output end and the thermoviscous losses due to wall friction, as given by Silva\cite{silva:2009}. The effect of tone holes is not considered in this work.


\subsection{Nonlinear coupling}
As described by Hirschberg\cite{hirschberg:1995} and further confirmed experimentally by Dalmont et al.\cite{dalmont:2003} and Almeida et al.\cite{almeida:2007a}, the flow inside the reed channel forms a jet that is dissipated by turbulence in the larger part of the mouthpiece leading to the following nonlinear equation:
\begin{equation}\label{eq:bernoulli}
U_j(t) = \mathrm{sign}(\Delta P) W h(t) \sqrt{\frac{2|\Delta P|}{\rho}}
\end{equation}
where $W$ is the width of the reed channel, assumed to be constant, and $U_j$ has the sign of $\Delta P$. According to this model, the direction of the flow depends on the sign of the pressure drop across the reed $\Delta P$.

\subsection{Reed motion induced flow}
Moreover, when an oscillation occurs (a note being played), the reed periodic motion induces an acoustic volume flow in addition to the one of the jet, as early described by Thompson\cite{thompson:1979} and lately studied by Silva et al.\cite{silva:2008}. Writing $S_r$ the reed effective area $S_r$, the resulting flow $U_r$ reads:
\begin{equation}\label{eq:reed_flow}
U_r(t)=-S_r \frac{dh}{dt}(t).
\end{equation}
Notice the negative sign, due to the chosen representation where the reed tip position $h$ is positive at rest and null when closing the reed channel.

Some authors\cite{dalmont:1995,kergomard:livre} use another notation, the equivalent length correction $\Delta L$, related to $S_r$ as:
$$\Delta L = \frac{\rho c^2}{K_a S} S_r.$$
We prefer to stick with the notation $S_r$ but we will give the corresponding values of $\Delta L$ for comparison, when necessary.

\subsection{Global flow}
The global flow entering the resonator then reads:
\begin{equation}\label{eq:global_flow}
U(t) = \mathrm{sign}(\Delta P) W h(t) \sqrt{\frac{2|\Delta P|}{\rho}} - S_r \frac{dh}{dt}
\end{equation}

\subsection{Dimensionless model}
The above-described multi-physics model involves several variables. Expressed in SI units, their respective values are of very different orders of magnitude. However, as we will use numerical tools to solve the algebro-differential system composed of eq. \ref{eq:reed}, \ref{eq:imped}, \ref{eq:ptot} and \ref{eq:global_flow}, it is a useful precaution to use a dimensionless model.

In the dimensionless model, each variable should be divided by a typical value, in order to scale the range of possible values for that variable as close as possible to $[0,1]$.

Choosing the static difference of pressure necessary to close the reed channel $P_M=K_ah_0$ to scale all pressures, and the reed tip opening at rest $h_0$ to scale the reed tip position, the following new dimensionless variables and parameter are defined:
\begin{align*}
x(t) &= h(t)/h_0\\
y(t) &= h'(t)/v_0\\
p_n(t) &= P_n(t)/P_M \;\;\text{   for } \; n = 1,2,...N_m\\
p(t) &= P(t)/P_M\\
u(t) &= U(t) Z_c/P_M
\end{align*}
where $v_0=h_0\wr$ is a characteristic velocity of the reed.

The system of algebraic and differential equations (\ref{eq:reed},\ref{eq:imped},\ref{eq:ptot},\ref{eq:global_flow}) can be rewritten as follows (time-dependence is omitted, for a more concise writing):
\begin{equation}\label{eq:system_adim}
\left\{
\begin{array}{r l r}
\frac{1}{\wr}x' &= y\\
\frac{1}{\wr}y' &= 1-x + p -\g -q_ry\\
p_n' &= C_n u + s_n p_n \;\;\; \text{  for  } \; n = 1, 2, ...N_m\\
p &= 2\sum\limits_{n=1}^{N_m} \Re\big(p_n\big)\\
u &= \mathrm{sign}(\g-p)\zeta x \sqrt{|\g-p|} - \frac{v_0 Z_c}{P_M} S_r y,
\end{array}\right.
\end{equation}
where $\g = P_m/P_M$ is the dimensionless blowing pressure and $\zeta=Z_cW\sqrt{2h_0/\rho K_a}$ is the dimensionless reed opening parameter. The reader will notice that the parameter $\zeta$, related to the maximum flow through the reed channel, mainly depends on the geometry of the mouthpiece, the reed mechanical properties, as well as the player's lip force and position on the reed that control the opening.

\vskip3mm\par In the following sections we will compute the minimal mouth pressure $\g_{th}$ necessary to initiate self-sustained oscillations from the static regime and show how this threshold is modified when a second parameter of the model varies. When players change the way they put the mouthpiece in their mouth, several parameters of our model are assumed to vary : the control parameter $\zeta$, the reed modal damping $q_r$, its modal angular frequency $\wr$, as well as the effective area $S_r$ of the reed participating to the additional flow. We will first investigate the influence of the dimensionless parameter $k_r L=\wr L /c$, comparing with previous works. Then the influence of $\zeta$, $q_r$ and $S_r$ will also be investigated.

\section{Methods : Theoretical principles and numerical tools\label{sec:cont}}
In this section, the method used is briefly reviewed from a theoretical point of view. Practical application of the method will be carried out in section \ref{sec:results}.

\subsection{Branch of static solutions}
We consider an autonomous nonlinear dynamical system of the form:
\begin{equation}\label{eq:ds}
\mathbf{u}'(t) = \mathbf{F}(\mathbf{u}(t),\l)
\end{equation}
where $\mathbf{u}(t)$ is the state vector and $\l$ is a chosen parameter of the system equations.

Let $\mathbf{u}_0 \in \mathbb{R}^n$ be a fixed point of this system for $\l_0$: $\mathbf{F}(\mathbf{u}_0,\l_0)=0$.
Under the hypothesis that $\mathbf{u}_0$ is a regular solution, there exists a unique function $\mathbf{u}(\l)$ that is solution of the previous equation for all $\l$ close to $\l_0$, with $\mathbf{u}(\l_0) = \mathbf{u}_0$. (See Doedel\cite{doedel:2010} for formal definitions, as well as for more details about regular solutions and what is meant by ``close''.)

In other words, for $\l$ close to $\l_0$, there is a continuum of equilibrium points that passes by $\mathbf{u}_0$ and the \emph{branch} of solutions is represented by the graph $u_i=f(\l)$, where $u_i$ is a component of the state vector $\mathbf{u}$.

\subsection{Continuation of static solutions\label{subsec:static-cont}}
Considering the differential equations in system (\ref{eq:system_adim}), it can be rewritten in the form $\mathbf{u}'=\mathbf{F}(\mathbf{u},\l)$ where $\mathbf{u}$ is the state vector and $\l$ is one of the equations parameters (for instance the blowing pressure $\g$). Thus, a static solution is given by the algebraic system: $\mathbf{F}(\mathbf{u},\l)=0$, where the algebraic equations of system (\ref{eq:system_adim}) can now be included.

The branch of static solutions is computed numerically using a classical path following technique based on Keller's pseudo-arc length continuation algorithm\cite{keller:1977} as implemented in the softwares AUTO (Doedel and Oldeman \cite{doedel:2009}) and MANLAB (Karkar et al. \cite{manlab}), both freely available online. For comprehensive details about continuation of static solutions using these two numerical tools, please refer to the conference article \cite{karkar:ica2010}.

\subsection{Hopf Bifurcation}
Let us consider a static solution $\mathbf{u}$ of system (\ref{eq:ds}). The stability of this equilibrium point is given by eigenvalues of the jacobian matrix of the system: $$\mathbf{F}_u(\mathbf{u},\l)= \left[\frac{\partial \mathbf{F}_i}{\partial \mathbf{u}_j}\right]$$. If all the eigenvalues of $\mathbf{F}_u$ have a strictly negative real part, then the equilibrium is stable. If any of its eigenvalues has a strictly positive real part, then the equilibrium is unstable.

Several scenarios of loss of stability, along the branch $\mathbf{u}=f(\l)$ are possible. One of them is the following : a unique pair of complex conjugate eigenvalues crosses the imaginary axis at $(i\w_{th},-i\w_{th})$ for $\l=\l_{th}$.

In that case, the system is said to undergo a \emph{Hopf Bifurcation}. It means that a family of periodic orbits starts from this point $\mathbf{u}(\l_{th})$, with an angular frequency $\w_{th}$. As for the clarinet model, choosing the blowing pressure $\g$ as the varying parameter, the first Hopf bifurcation encountered when $\g$ is increased from 0 is the oscillation threshold: it is the minimal blowing pressure $\g_{th}$ needed to initiate self-sustained oscillations (i.e. to play a note) when starting from zero and increasing quasi-statically the blowing pressure.

\subsection{Branch and continuation of Hopf bifurcations}
This definition of a Hopf bifurcation can be written as an extended algebraic system $\mathbf{G}=0$ where $\mathbf{G}$ reads:
\begin{equation}\label{eq:HB}
\left\{
\begin{array}{l}
\mathbf{F}(\mathbf{u},\l,\mu)\\
\mathbf{F}_{\mathbf{u}}(\mathbf{u},\l,\mu) \mathbf{\phi} - j \w \mathbf{\phi}\\
\mathbf{\phi}^T\mathbf{\phi}-1
\end{array}\right.
\end{equation}
$\l$ and $\mu$ are two parameters of interest (for instance the blowing pressure $\g$ and the reed opening parameter $\zeta$), and $\mathbf{\phi}$ is the normed eigenvector associated with the purely imaginary eigenvalue $j \w$.

Assuming that a given solution $X_0=(u_0,\phi_0,\w_0,\l_0,\mu_0)$ is a regular solution of $\mathbf{G}=0$, there exists a continuum of solutions $X(\mu)=(u(\mu),\phi(\mu),\w(\mu),\l(\mu),\mu)$ near $X_0$. Thus, the function $X(\mu)$ is a branch of Hopf bifurcations of our initial dynamical system.

Using the same continuation techniques as in \ref{subsec:static-cont}, this branch of Hopf bifurcations can be computed. The reader is kindly referred to classical literature on the subject for proper theoretical definitions and other details about the continuation of Hopf bifurcations (see for instance Doedel\cite{doedel:2010}).

\section{Results\label{sec:results}}
\begin{figure}
\includegraphics[width=\columnwidth]{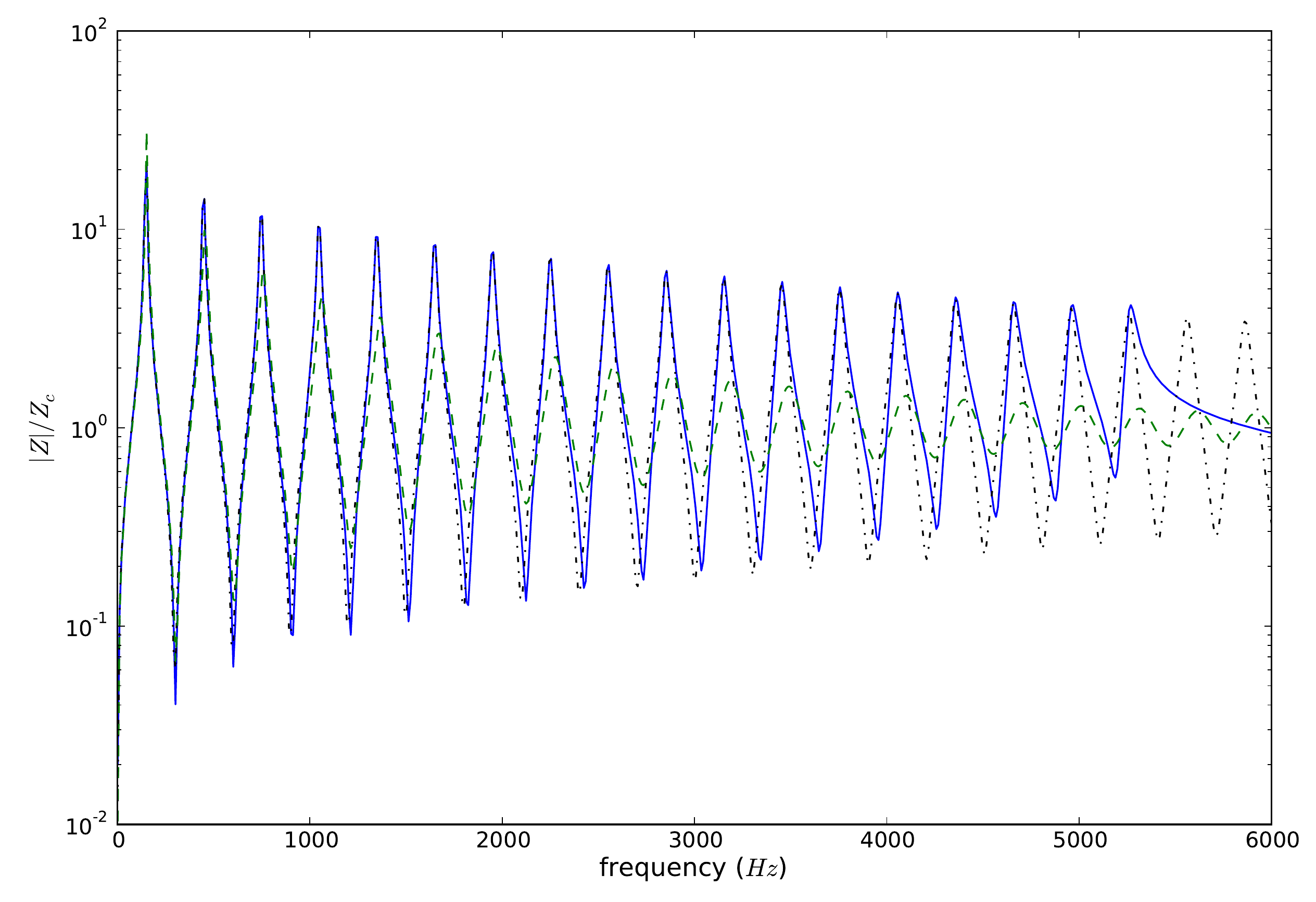}
\caption{\small Impedance spectrum (reduced modulus) of a purely cylindrical bore of length $L$=$57$cm taking into account thermoviscous losses and radiation and truncated to 18 modes as used in this study (---), analytical impedance spectrum without modal decomposition (-- $\cdot$ --), impedance spectrum used in Wilson and Beavers\cite{wilson:1974}. (-- --).\label{fig:impedance}}
\end{figure}

In the current section, the method described in the previous section is applied to the physical model presented in section \ref{sec:model} to investigate the oscillation threshold of a purely cylindrical clarinet. The resonator modal decomposition has been computed with the MOREESC\cite{moreesc} software, using 18 modes. The corresponding input impedance spectrum is shown on figure \ref{fig:impedance}. For comparison, the impedance spectrum given by the analytical formulae of Backus\cite{backus:1963} and used in Wilson and Beavers\cite{wilson:1974} is also plotted.
Important differences appear concerning frequency dependent damping (peaks heights) and harmonicity (peaks positions).

\subsection{Reed-bore interaction}
\subsubsection{Comparison with previous results}
\begin{figure}
\includegraphics[width=\columnwidth]{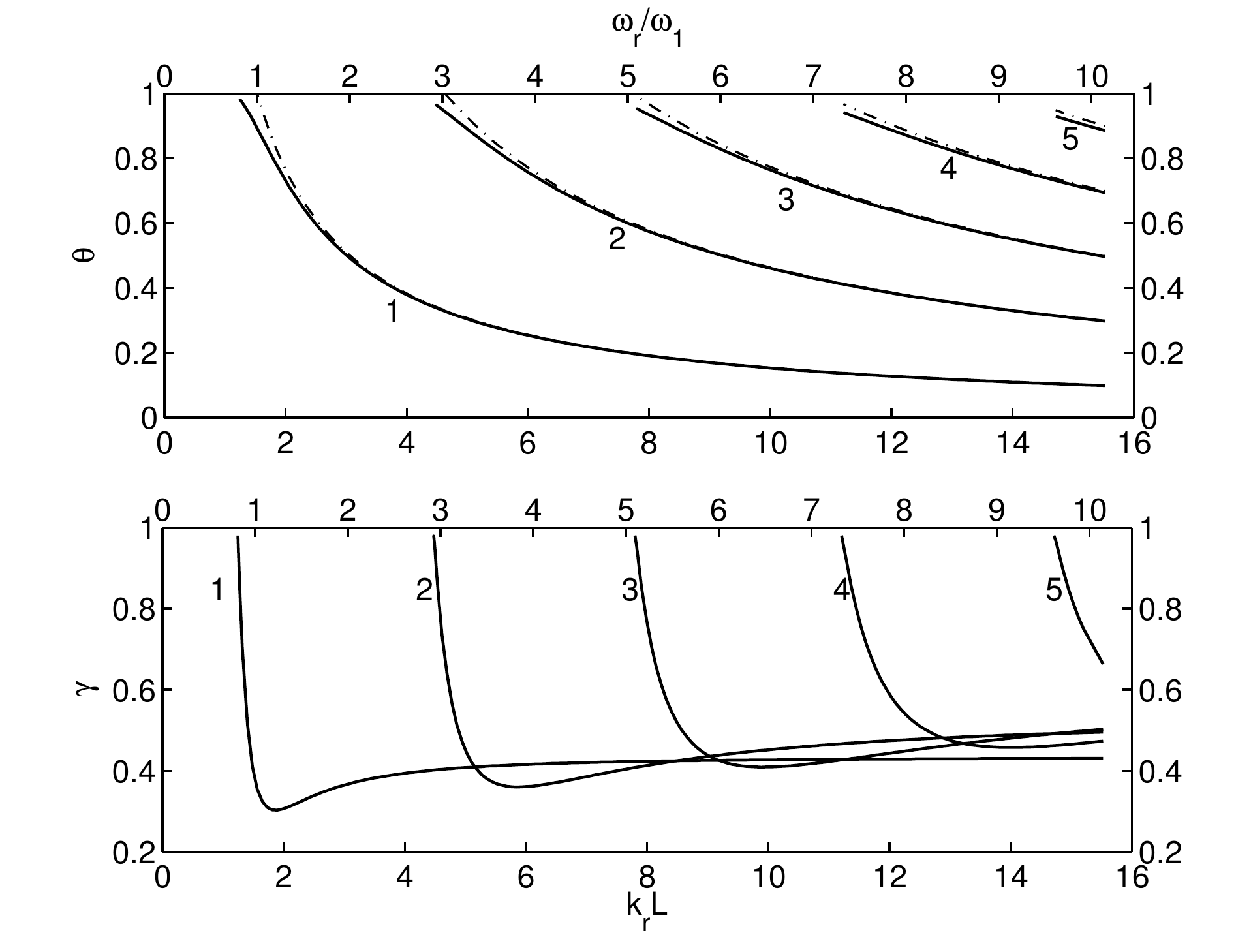}
\caption{\label{fig:suivi_krL.eps}\small Critical blowing pressure and frequencie of each Hopf bifurcation (labelled from 1 to 5) as a function of the ``tube parameter'' $k_rL$ (as used in W\&B), or the frequency ratio $\wr/\w_1$ (secondary x-axis on top). Lower plot: critical blowing pressure $\g$. Upper plot: frequency ratio $\theta=\w/\wr$ of the oscillations; passive resonances of the bore $\theta_n=\w_n/\wr$ are plotted for comparison (-- $\cdot$ --). Physical constants and parameters of the model were chosen according to fig. 1 in Silva et al.\cite{silva:2008}: $\rho$=$1.185$g/L, $c$=$346$m/s, $L$=$57$cm, $r$=$7$mm, $q_r$=$0.4$, $\zeta$=$0.13$, $S_r$=$0$, varying $\wr$.}
\end{figure}
Previous works of Wilson and Beavers\cite{wilson:1974} and Silva et al.\cite{silva:2008} showed how the resonance of the reed competes with the acoustical modes of the resonator for the existence of self-sustained oscillations: oscillation thresholds and emergent frequencies were then measured and numerically computed, for different values of the dimensionless product $k_rL$ ($k_r=\wr/c$ is the corresponding wavenumber of the reed modal frequency, and $L$ is the length of the bore --which is approximately the quarter of the first acoustical mode wavelength) called ``tube parameter'' in W\&B. Notice that in these studies, the parameter $k_r$ was kept constant and $L$ was varied. These results showed that several regimes can be selected, depending on the product $k_rL$, only if the reed damping is very small.
\begin{figure}
\includegraphics[width=\columnwidth]{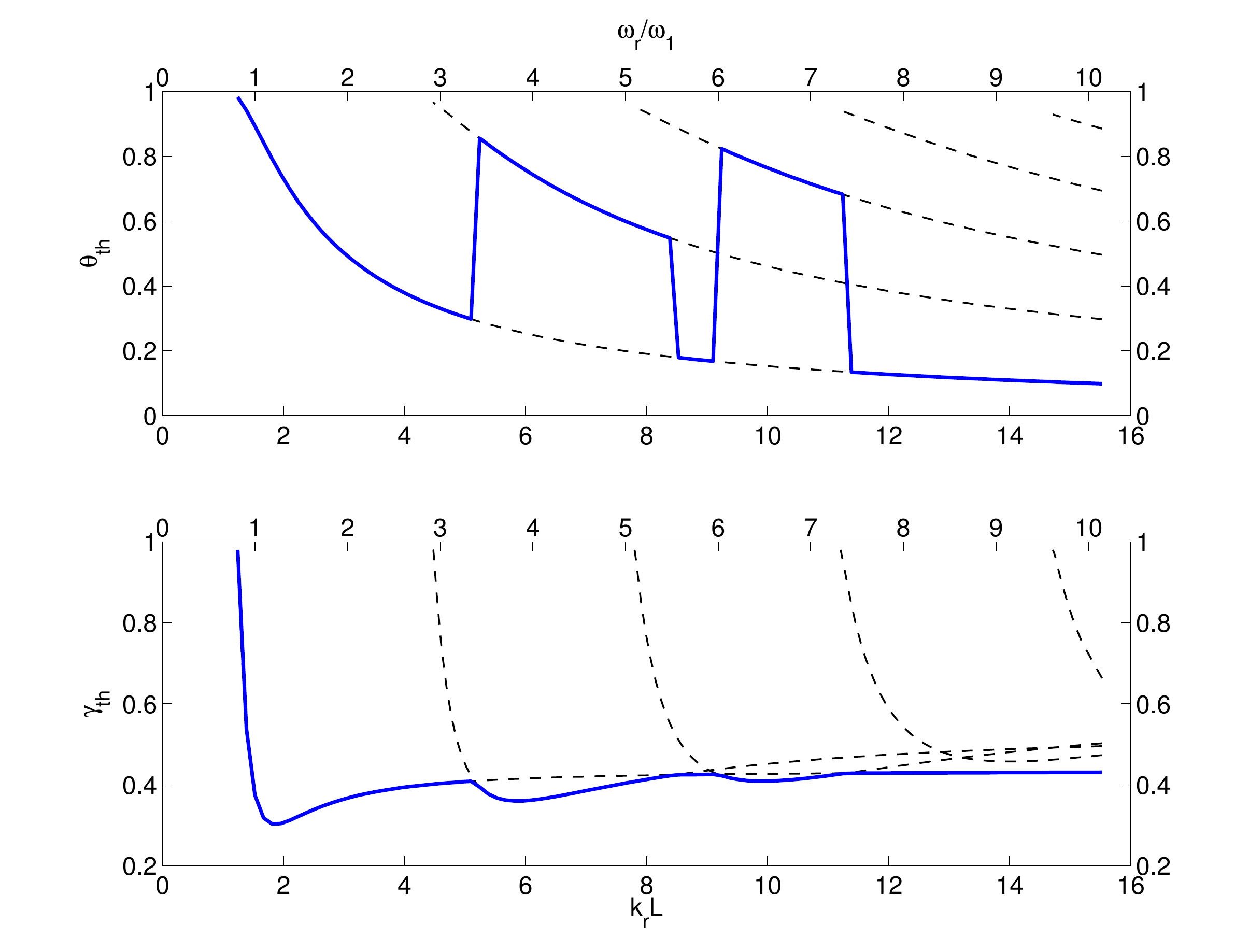}
\caption{\small Oscillation threshold : blowing pressure, regime selection and frequency with respect to $k_rL$. The oscillation threshold, for a given abscissa, is given by the lowest blowing pressure of the five Hopf bifurcation branches (thick line). Lower plot: blowing pressure threshold $\g_{th}$. Upper plot: frequency ratio $\theta_{th}=\w_{th}/\wr$ of the oscillations. Complete branches of the five bifurcations are reminded (dashed lines). Same model parameters were used as fig. \ref{fig:suivi_krL.eps}. A secondary x-axis on top of each plot gives the value of the frequency ratio $\wr/\w_1$.
\label{fig:suivi_krL_min.eps}}
\end{figure}

Figure \ref{fig:suivi_krL.eps} shows similar computations using our method, with the same model parameters as was used for the figure 1 of Silva et al.\cite{silva:2008}, only using a different impedance (as explained above). A secondary x-axis, on top of each plot, shows the corresponding ratio of the reed natural frequency $\wr$ to the first resonance frequency of the bore $\w_1=\Im(s_1)$. Note that instead of varying $L$ for a given $\wr$, the figure was computed by varying $\wr$ for a given $L$. The method advantage, here, is that no analytical development of the impedance spectrum is needed, unlike what was done by W\&B to derive the characteristic equation. Thus, any other impedance could be used, for instance a measured one.

Figure \ref{fig:suivi_krL_min.eps} illustrates how the oscillation threshold is deduced from the previous figure: for a given abscissa, it corresponds to the lowest of the five Hopf bifurcation branches, each corresponding to a register. In contrast with W\&B, this figure clearly shows that even in the case of a strongly damped reed ($q_r=0.4$), a regime selection occurs with varying $w_r$. The term ``strongly damped'' was used by W\&B for that value of $q_r$, but it seems to be a realistic value for a clarinet reed coupled to the player's lip (see van Walstijn and Avanzini\cite{vanwalstijn:2004} and Gazengel et al.\cite{gazengel:2007} for numerical and experimental studies of the reed-lip system). A minimum of the threshold blowing pressure (lower plot) appears a little above $\wr=\w_1$, i.e. when the first air column resonance frequency is close to the reed one. The frequency of a given mode (upper plot) is close to the corresponding passive resonance frequency $\w_n$=$\Im(s_n)$, for medium and high values of $k_rL$, but tends to the reed modal frequency for low values of $k_rL$.

\subsubsection{Relevance of the tube parameter $k_rL$}
While results of the previous figure are in good agreement with Silva's results (when thermoviscous losses are taken into account, as in figure 3 of Silva et al.\cite{silva:2008}), several differences lead to question the relevance of the representation: if $k_rL$ is a characteristic parameter of the model, varying $L$ for a given $\wr$ should be equivalent to varying $\wr$ with a fixed value of $L$.

\begin{figure}
\includegraphics[width=\columnwidth]{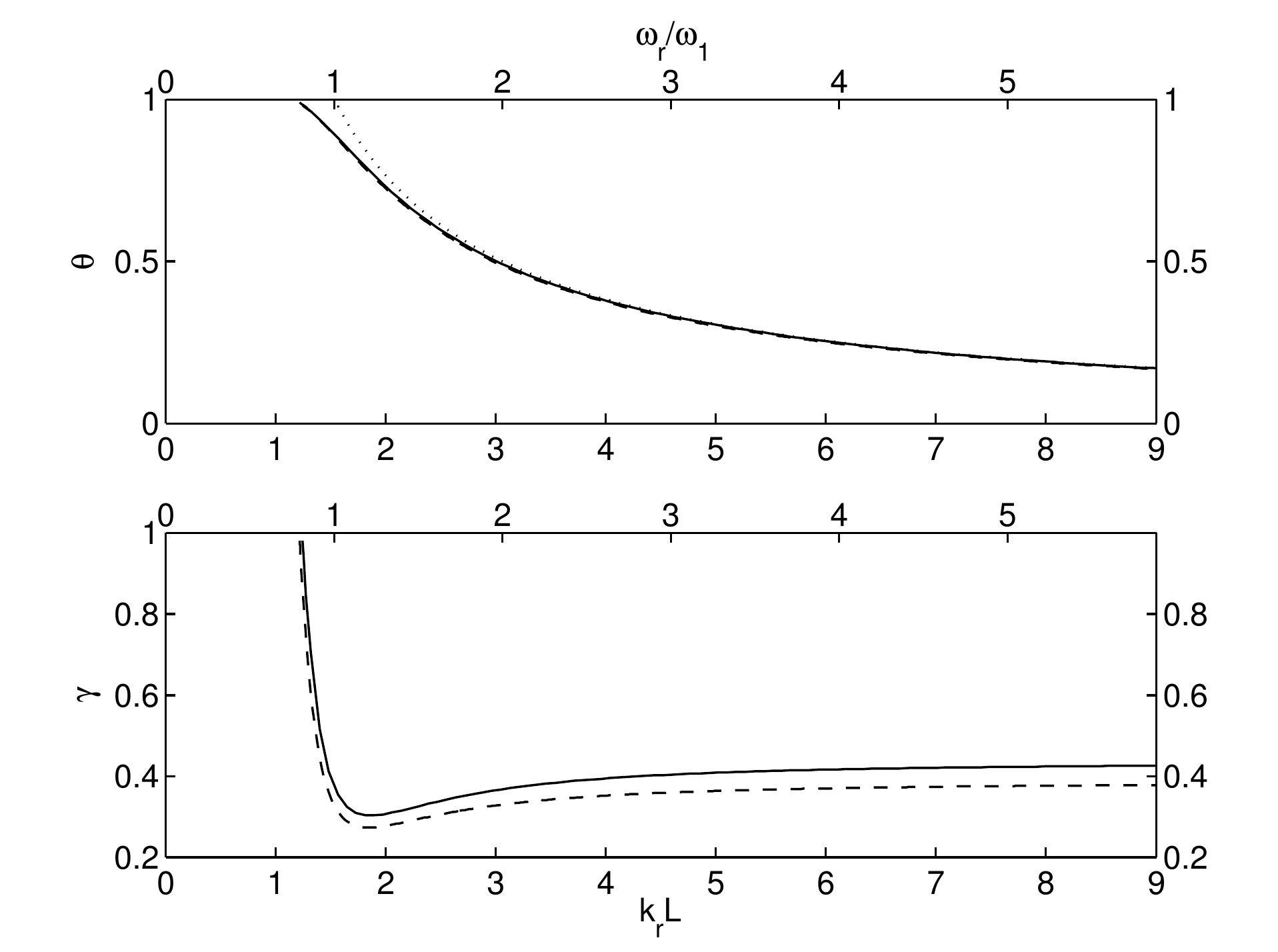}
\caption{\small Comparison of the first regime for $L$=$57$cm (---) and
$L$=$14.69$cm (-- --). Lower plot: dimensionless blowing pressure $\g$ of the bifurcation. Upper plot: corresponding dimensionless frequency ratio $\theta=\w/\wr$; the first passive resonance of the bore is reminded ($\cdots$).
 \label{fig:suivi_krL_comparaison_L}}
\end{figure}

To answer this question, we recomputed the first regime for the same values of $k_rL$ but with $L$=$14.69$cm (one then needs to recompute the modal coefficients of the input impedance). The results are plotted in figure \ref{fig:suivi_krL_comparaison_L}: considering the upper plot, the frequency seems to be independent of $L$, as long as the product $k_rL$ is kept constant (the plain line and dashed line are exactly superimposed); however, considering the lower plot, the blowing pressure does not behave in the same way, for identical values of $k_rL$, depending on the value of $L$.
Therefore, the tube parameter $k_rL$ is not a characteristic invariant parameter of the model.

One good reason for that is that in our impedance spectrum, the peaks are inharmonic and have a frequency dependent quality factor and magnitude. Then varying $L$ does not preserve the impedance peaks height and width, nor their spacing, which leads to a different balance in the competition with the reed resonance.

\subsection{Simultaneous influence of reed damping and modal frequency}
In previous works\cite{wilson:1974,silva:2009}, figures similar to figure \ref{fig:suivi_krL.eps} were plotted for two cases, small and large values of the reed damping $q_r$, which revealed very different behaviours. 

\begin{figure}
\includegraphics[width=\columnwidth]{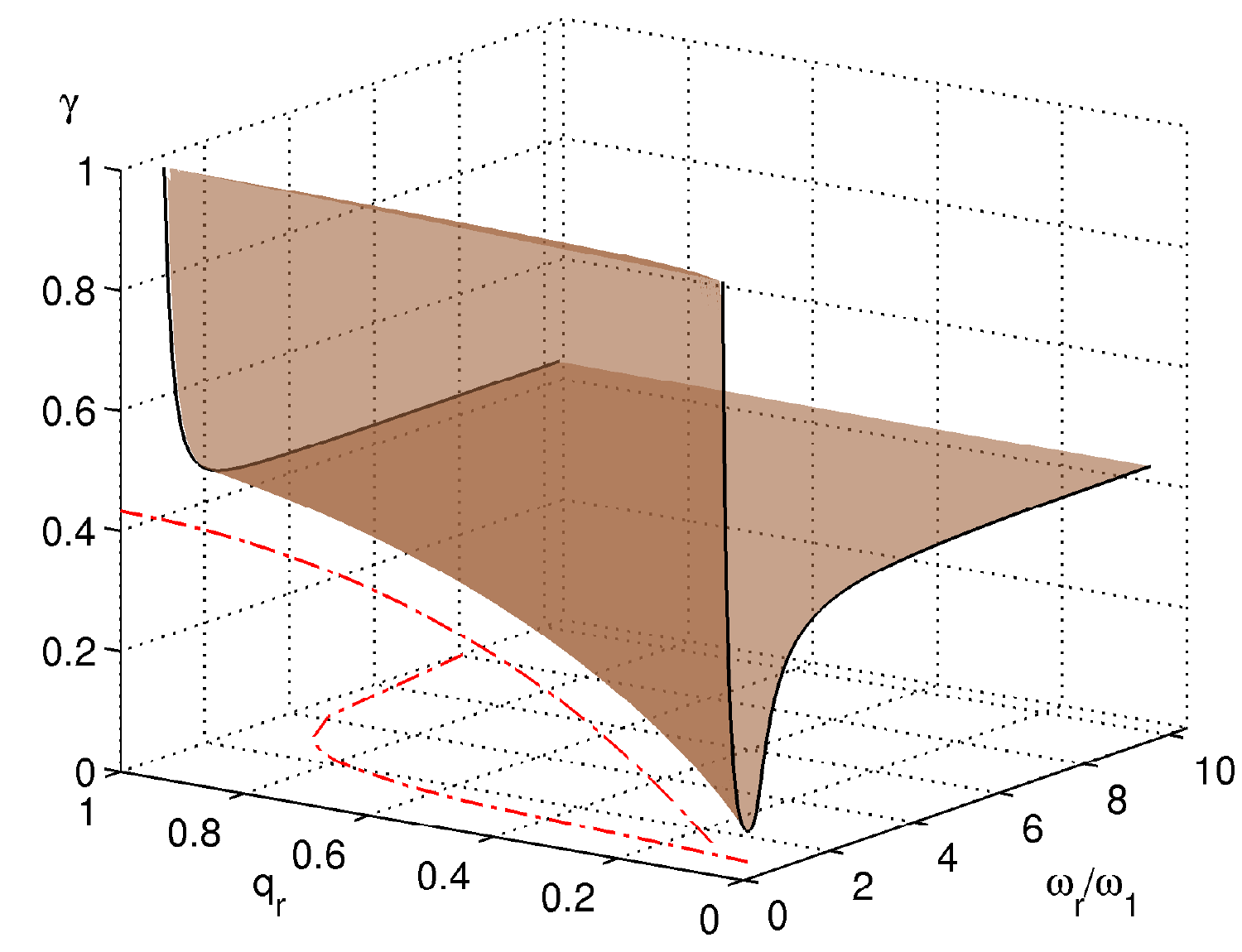}
\caption{\small Surface giving the critical blowing pressure $\gamma$ of the Hopf bifurcation corresponding to the first register, with respect to the reed damping parameter $q_r$ and the reed modal frequency $\wr$ (divided by the first bore resonance $\w_1$). The plain curves correspond to the limits of the chosen range for $q_r$ , i.e. $q_r=0.05$ and $q_r=1$. The dot-dashed curve in the plane $\w_r/\w_1=0$ corresponds to the minimum $\gamma_0$, and the dot-dashed curve in the plane $\gamma=0$ corresponds to the value of $\w_r/\w_1$ at which it occurs.
\label{fig:3D_gamma_wr_qr_1stReg}}
\end{figure}

Because the method proposed in this paper leads to very short computation time, it is quite easy to loop the computation for a series of $q_r$ values. For sufficiently close values of $q_r$, this allows to draw a tridimensional plot in which the surface represents the critical value of the dimensionless blowing pressure $\g$ corresponding to a given Hopf bifurcation as a function of two parameters, the reed damping $q_r$ and the frequency ratio $\wr/\w_1$, as illustrated figure \ref{fig:3D_gamma_wr_qr_1stReg} for the Hopf bifurcation of the first register. The plotted surface corresponds to a Hopf bifurcation locus when two parameters are varied (conversely to figure \ref{fig:suivi_krL.eps} where only one parameter is varied). The two plain, thick curves correspond to the limits of the domain considered: $q_r=0.05$ and $q_r=1$.

This figure allows to investigate the transition between large and light damping.
It appears that for large values of $\wr/\w_1$, the critical blowing pressure is nearly independent of $q_r$: a closer look reveals a slight increase with $q_r$. This is not surprising since $\wr/\w_1>>1$ corresponds to the case where the first resonance frequency of the resonator is very small compared to $\omega_r$. Hence, only very large values of $q_r$ ($q_r>>1$) would contribute significantly to an increase of the pressure threshold. However, for smaller values of $\wr/\w_1$, the value of $q_r$ becomes determinant.	 In the 2D manifold, there is a valley which becomes deeper when $q_r$ decreases. For a given $q_r$, the bottom point of the valley (called $\gamma_0$ in Silva et al.\cite{silva:2008}) corresponds to the minimum threshold. It is plotted as a dot-dashed line on the plane ($\wr/\w_1=0$), whereas the corresponding abscissa $\left.\frac{\wr}{\w_1}\right|_0$ is plotted as a dot-dashed line in the plane $\gamma=0$. It shows that the minimum value $\gamma_0$ is an increasing function of $q_r$. The same conclusion holds for $\left.\frac{\wr}{\w_1}\right|_0$.

Notice that the chosen range for the values of $q_r$ seems quite realistic, according to literature on the subject: in a numerical model, van Walstijn and Avanzini\cite{vanwalstijn:2004} reported parameters values equivalent to $q_r=0.24$ for one playing condition, whereas Gazengel et al.\cite{gazengel:2007} found experimental values going from $q_r=0.05$ for the bare reed to $q_r=1.54$ for a high lip pressure on the reed. Thus, the lower value $q_r=0.05$ corresponds to the limit case of a very resonant reed with no lip pressing on it, and the higher value $q_r=1.00$ is high enough to cover a fairly good range of lip pressures.

\begin{figure}
\includegraphics[width=\columnwidth]{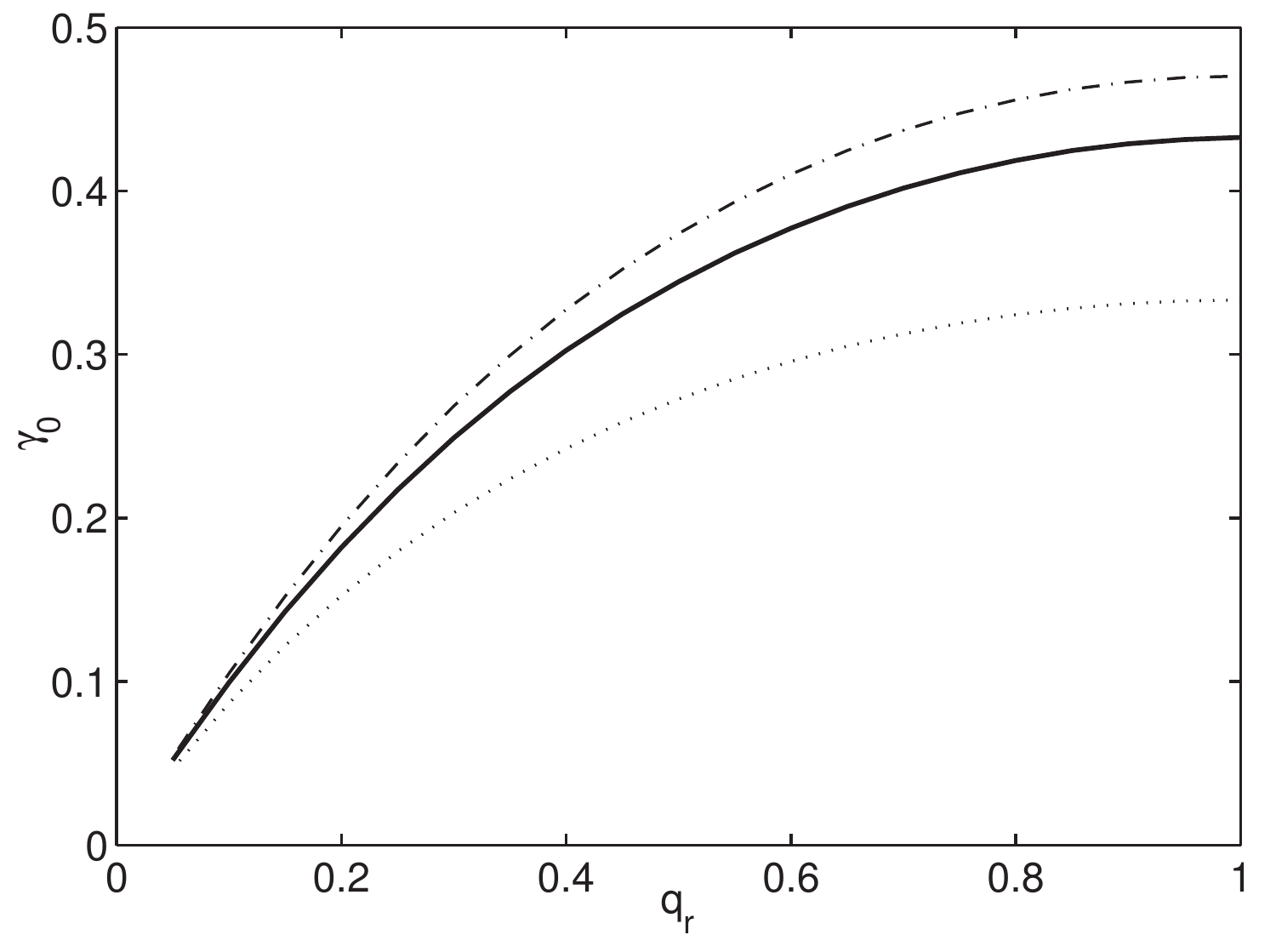}
\caption{\small Minimum threshold $\g_0$ as a function of $q_r$. Our numerical results (---), without approximation. Analytical results using two approximations: no loss ($\cdots$), single-mode resonator with losses (-- $\cdot$ --).\label{fig:gamma_min_compare_analytic}}
\end{figure}
Now we illustrate that the results obtained on this model allow to estimate the range of validity of analytical formula obtained in approximated cases. For instance, in figure \ref{fig:gamma_min_compare_analytic}, the minimum $\gamma_0$ obtained through numerical continuation (without approximation), is compared with two analytical formula from Silva\cite{silva:2009} corresponding to different approximations: no losses in the cylindrical bore (here plotted with dotted line, corresponds to eq. (14) in Silva et al.\cite{silva:2008}) and a single-mode resonator with losses (dot-dashed line, eq. (19) in Silva et al.\cite{silva:2008}).

It appears that only one mode with viscothermal losses leads to a more precise result than the undamped formulation with all the modes. However, the relative deviation from our results still reaches $10\%$.

\begin{figure}
\begin{minipage}{\columnwidth}
\includegraphics[width=\columnwidth]{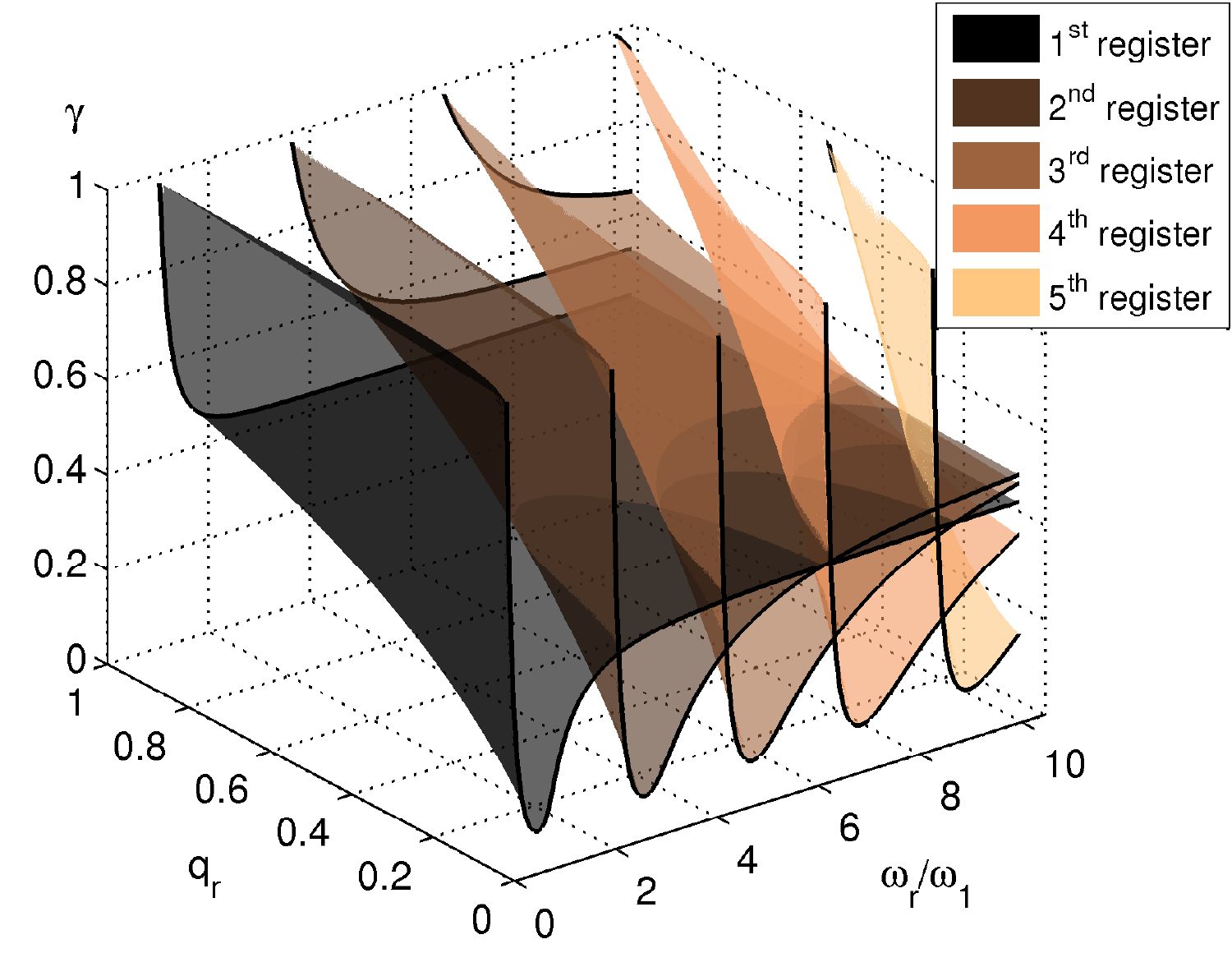}
\caption{\small a) Similar plot as in figure \ref{fig:3D_gamma_wr_qr_1stReg} for the first five registers: the critical blowing pressure $\gamma$ of each Hopf bifurcation is  plotted with respect to  the reed damping parameter $q_r$ and frequency ratio $\wr/\w_1$. The plain curves correspond to the smallest and highest $q_r$ where a bifurcation occurs on the chosen range.\label{fig:3D_gamma_wr_qr}}
\addtocounter{figure}{-1}
\includegraphics[width=\columnwidth]{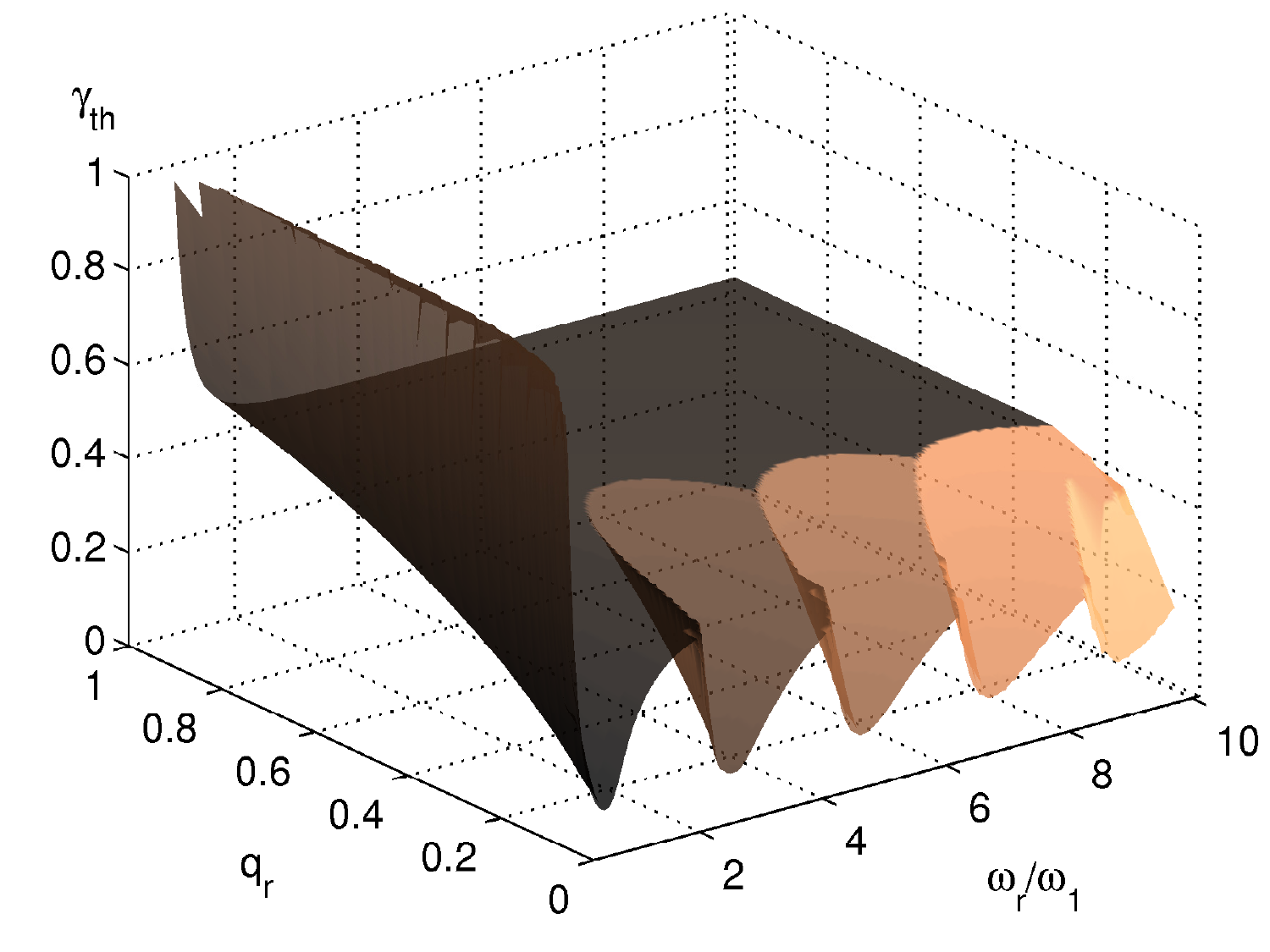}
\caption{\small b) The oscillation threshold extracted from figure \ref{fig:3D_gamma_wr_qr}-a: blowing pressure at threshold $\gamma_{th}$, defined by the lowest of the five surfaces plotted in figure 7-a.\label{fig:3D_gamma_wr_qr_min}}
\end{minipage}
\end{figure}
However, the oscillation threshold is not always given by the Hopf bifurcation corresponding to the first register. Considering not only the first register, but the first five registers, leads to figure \ref{fig:3D_gamma_wr_qr}-a. Note that the number of registers is not arbitrary: at the right end of the figure, where $\wr/\w_1\simeq10$ ($k_rL\simeq16$), the sixth resonance ($\w_6\simeq11\w_1$) is well beyond the reed resonance and thus cannot set up self-sustained oscillations because its Hopf bifurcation is beyond $\g>1$. More registers should be considered if greater $\wr$ values were to be explored. For sake of clarity, $\gamma_0$ and $\left.\frac{\wr}{\w_1}\right|_0$ are not plotted.
The plain lines in the plane $(q_r=0.05)$ shows similar behaviour as in figure \ref{fig:suivi_krL.eps} and has already been discussed. The other black plain lines depict a very different situation where the first register always correspond to the lowest $\gamma$.

Figure \ref{fig:3D_gamma_wr_qr}-b shows the oscillation threshold, as defined by the first Hopf bifurcation encountered when increasing the blowing pressure $\g$ from $0$, as a function of the reed damping $q_r$ and the frequency ratio $\wr/\w_1$. It is deduced from the previous one the same way figure \ref{fig:suivi_krL_min.eps} was deduced from figure \ref{fig:suivi_krL.eps}. Intersections between the five surfaces result in an oscillation threshold surface with several local minima.

It clearly shows how the reed damping, which can be control by the player with his lower lip, plays a key role in the register selection, as previously reported by W\&B\cite{wilson:1974} and Silva et al.\cite{silva:2008}. It is also clearly visible that the range of $q_r$ for which a given register exists decreases with the index of this register. For instance, considering the fifth register, it can only be selected for $q_r<0.3$.

\begin{figure}
\includegraphics[width=\columnwidth]{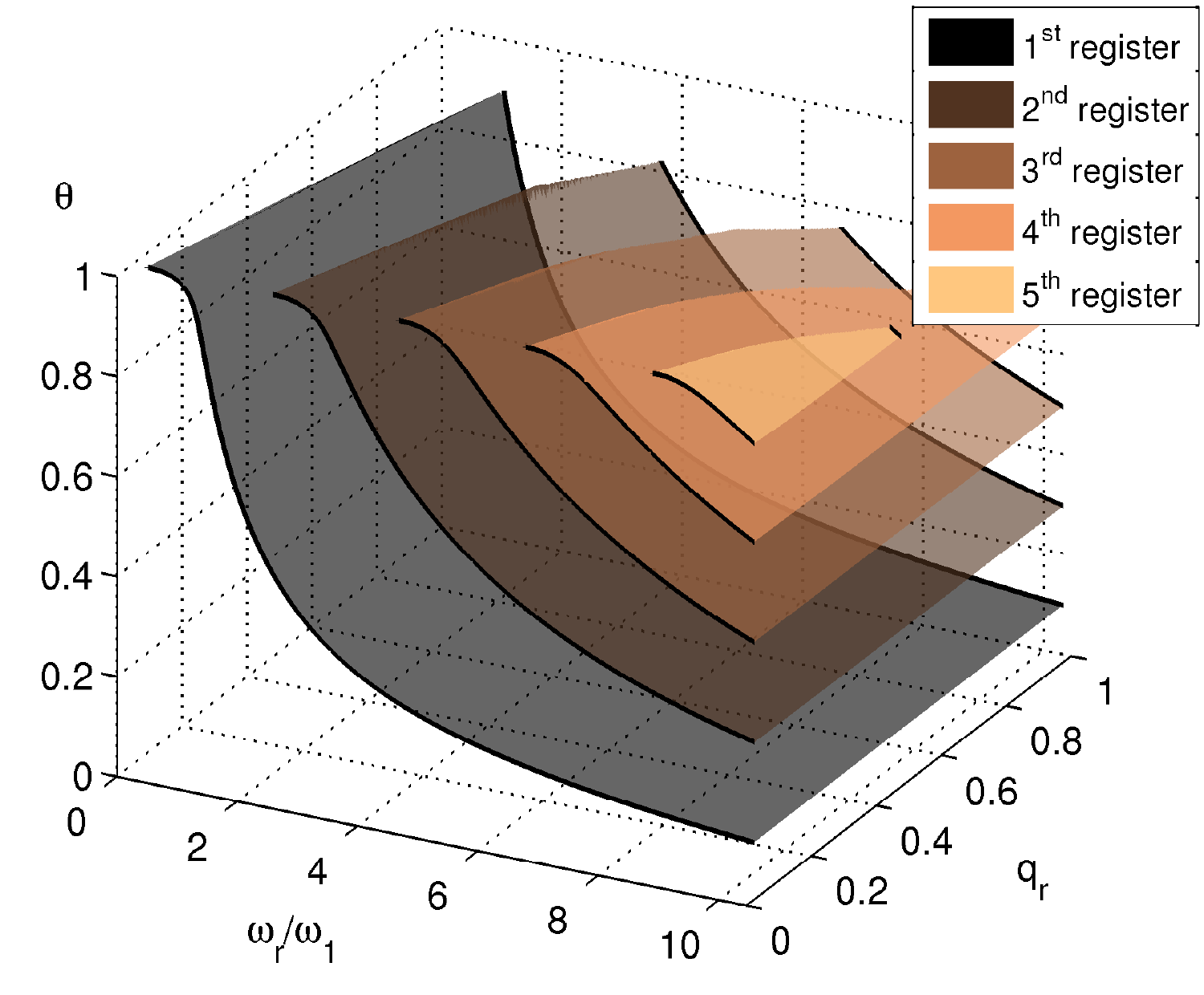}
\caption{\small Dimensionless frequency $\theta=\w/\wr$ at the bifurcation for the first five registers, with respect to the reed damping $q_r$ and the frequency ratio $\w_r/\w_1$.\label{fig:3D_theta_wr_qr}}
\end{figure}
The frequencies corresponding to the different registers are also calculated and plotted in figure \ref{fig:3D_theta_wr_qr}. The frequency at threshold appears to be an increasing function of $q_r$, but the influence of $q_r$ does not look significant in this 3D representation. When $q_r$ goes from 0 to 1, the typical relative frequency deviations from the bore resonances are less than $1\%$. However, a frequency shift more than $4\%$ can be observed for the first register around $\wr/\w_1=1.1$. Such a ratio is quite unusual for a clarinet, but it is of great interest for reed organ-pipe manufacturer, where the reed natural frequency is close to the bore resonance, and the damping very small.

\subsection{Influence of the control parameter $\zeta$}
Let us remind here the definition of this parameter: $\zeta=WZ_c\sqrt{2h_0/\rho K_a}$. Its variations are mainly related to the changes of the reed opening parameter $h_0$ during the play. Those are directly driven by changes of the player's lip pressure and position on the reed. Thus it is a very important control parameter of the model. Notice that players modify both $q_r$ and $\zeta$ when changing the embouchure.

\begin{figure}
\includegraphics[width=\columnwidth]{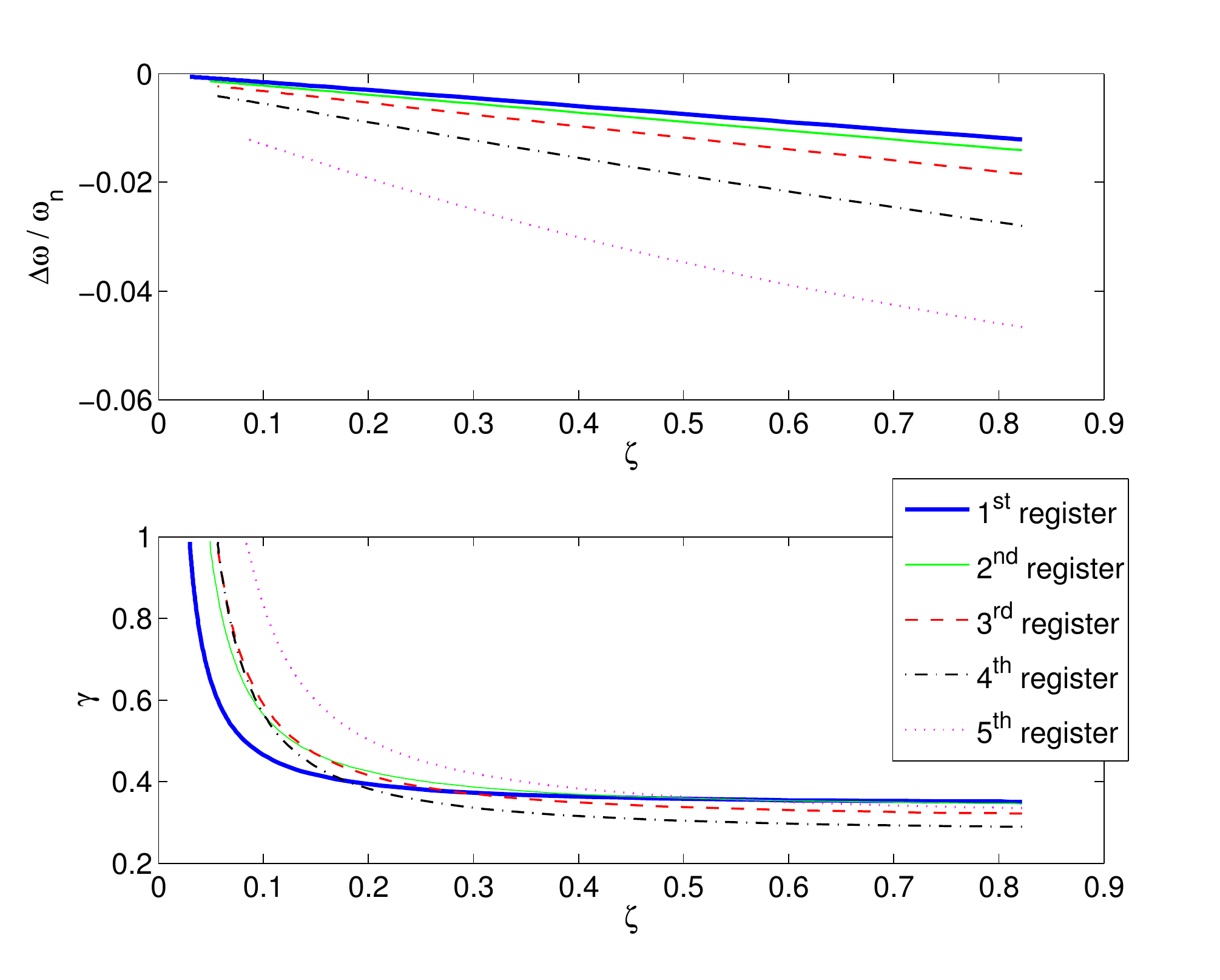}
\caption{\small Variations of the oscillation threshold with respect to $\zeta$. Lower plot: critical, dimensionless blowing pressure $\g$; the lowest curve defines the oscillation threshold $\g_{th}$ at a given abscissa. Upper plot: relative frequency deviation $\Delta\w/\w_n=(\w-\w_n)/\w_n$ between the frequency at threshold $\w_{th}$ and the corresponding acoustical mode frequency $\w_n=\Im(s_n)$. Model parameters: $q_r$=$0.4$, $fr$=$1500$Hz, $S_r$=$0$cm\tss{2}.\label{fig:2D_ThetaGamma_versus_zeta}}
\end{figure}

Figure \ref{fig:2D_ThetaGamma_versus_zeta} shows the influence of $\zeta$ on the blowing pressure and frequency at the oscillation threshold (expressed as the relative deviation to the corresponding bore resonance frequency). This control parameter happens to be critical for the regime selection : from very low values up to $\zeta$=$0.17$, the first regime (with a frequency close to the first acoustical mode of the bore $f_0$) is selected, while the fourth regime ($f\simeq7f_0$) is selected for higher values.

The control parameter $\zeta$ also has noticeable influence on the frequency of the oscillations at threshold: whereas the first regime frequency deviation is less than $0.3\%$ (5 cents) in the range of $\zeta$ where it is the selected regime, the fourth regime frequency deviation is as high as $2.7\%$ (46 cents) for $\zeta$=$0.8$. This is a very important feature that has been highlighted in a paper by Guillemain et al.\cite{guillemain:2010}, where the lip stress on the mouthpiece of a saxophone were measured while a player was playing, showing an adjustment of the lip stress on the reed in order to correct the tone shortly after the beginning of the oscillations. It could also explain the difficult reproducibility of measurements when fitting a clarinet or saxophone mouthpiece in an artificial mouth.

The frequency deviations are monotoneous, decreasing functions of $\zeta$, almost linear on the range of interest. Also, when $\zeta$ tends towards $0$, the frequency at threshold tends to the corresponding bore passive resonance frequency. This result was to expect, since the boundary condition at the input end tends to a Neumann condition (infinite impedance). However, for very low values of $\zeta$, the blowing pressure threshold $\g_{th}$ quickly increases, and eventually reaches $1$ which is the also the static closing threshold. In the case where $\g_{th}>1$, the reed channel is always closed and no sound is possible.

\subsection{Concurrent influence of $q_r$ and $\zeta$}
\begin{figure}
\includegraphics[width=\columnwidth]{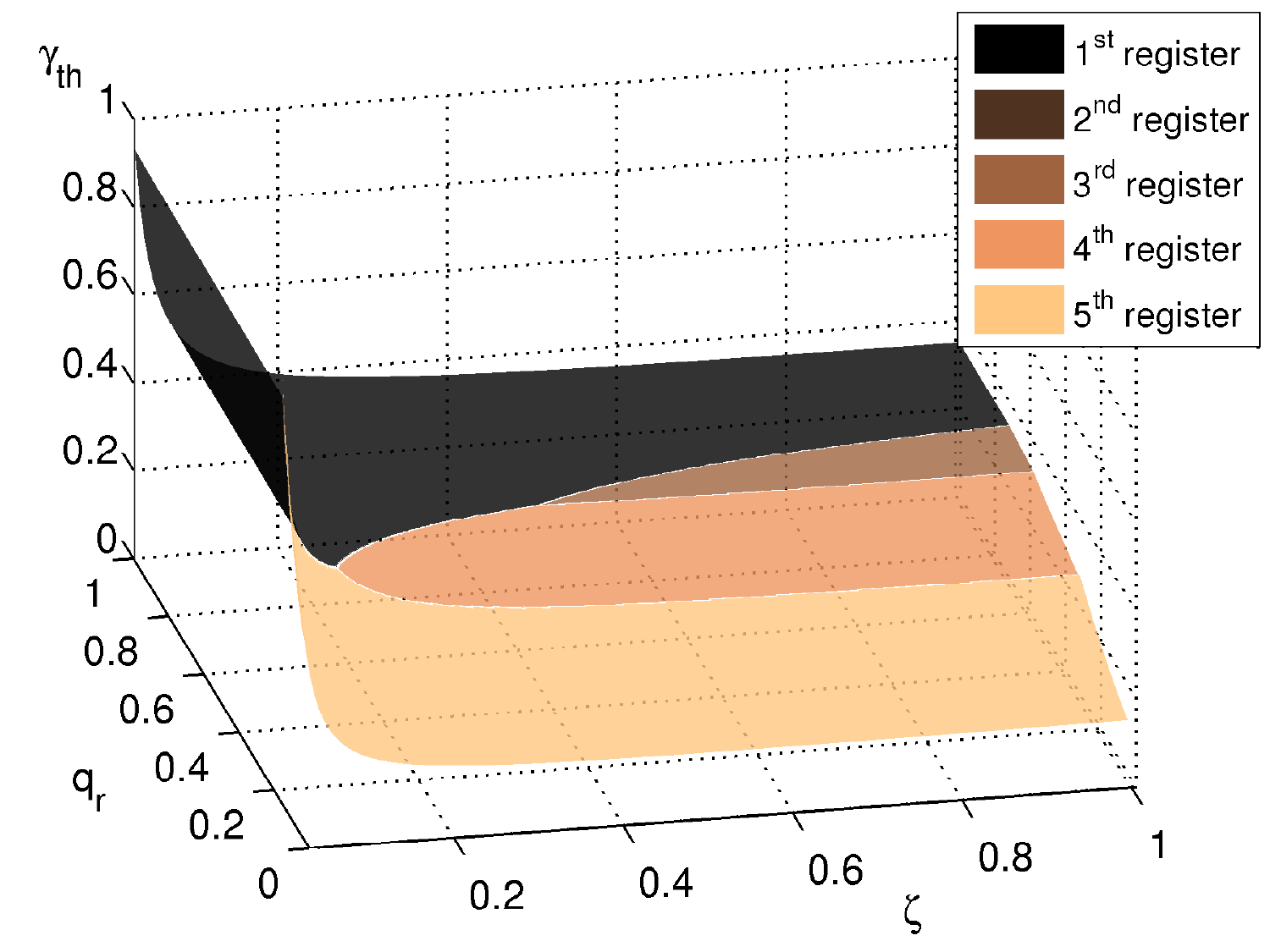}
\caption{\small Oscillation threshold with respect to the reed damping parameter $q_r$ and the control parameter $\zeta$. The blowing pressure threshold $\gamma_{th}$ is calculated the same way as in figure \ref{fig:3D_gamma_wr_qr}-b. The surface shading indicate the selected regime. \label{fig:3D_gamma_zeta_qr}}
\end{figure}

Figure \ref{fig:3D_gamma_zeta_qr} shows the blowing pressure threshold with respect to the reed damping $q_r$ and to the control parameter $\zeta$. For a given pair $(q_r,\zeta)$, the oscillation threshold is given by the lowest surface among the five 2D manifolds corresponding to the different registers.

Because the player's lower lip pressure and position on the reed, induce variations of both $\zeta$ and $q_r$ at the same time (the lip pressure being positively correlated with $q_r$ and negatively with $\zeta$), it is interesting to follow the oscillation threshold along the horizontal plane $\zeta=1-q_r$. On the one hand, for small values of $\zeta$ and $q_r$ close to unity (i.e. a high lip pressure) the first regime is clearly selected; on the other hand, for a high value of $\zeta$ and small damping $q_r$ (i.e. a relaxed lip on the reed), it is the fifth regime that is clearly selected. In between, all regimes are successively selected except the second regime that is noticeably absent of the figure.

\subsection{Influence of the reed motion induced flow\label{sec:reed_flow}}
In the numerical results presented so far, the reed induced flow was not taken into account ($S_r=0$ in all previous figures). However, as pointed out by Silva et al.~\cite{silva:2008}, it is important to take into account this additional flow, in order to predict accurately the emergent frequency at oscillation threshold.

\begin{figure}
\includegraphics[width=\columnwidth]{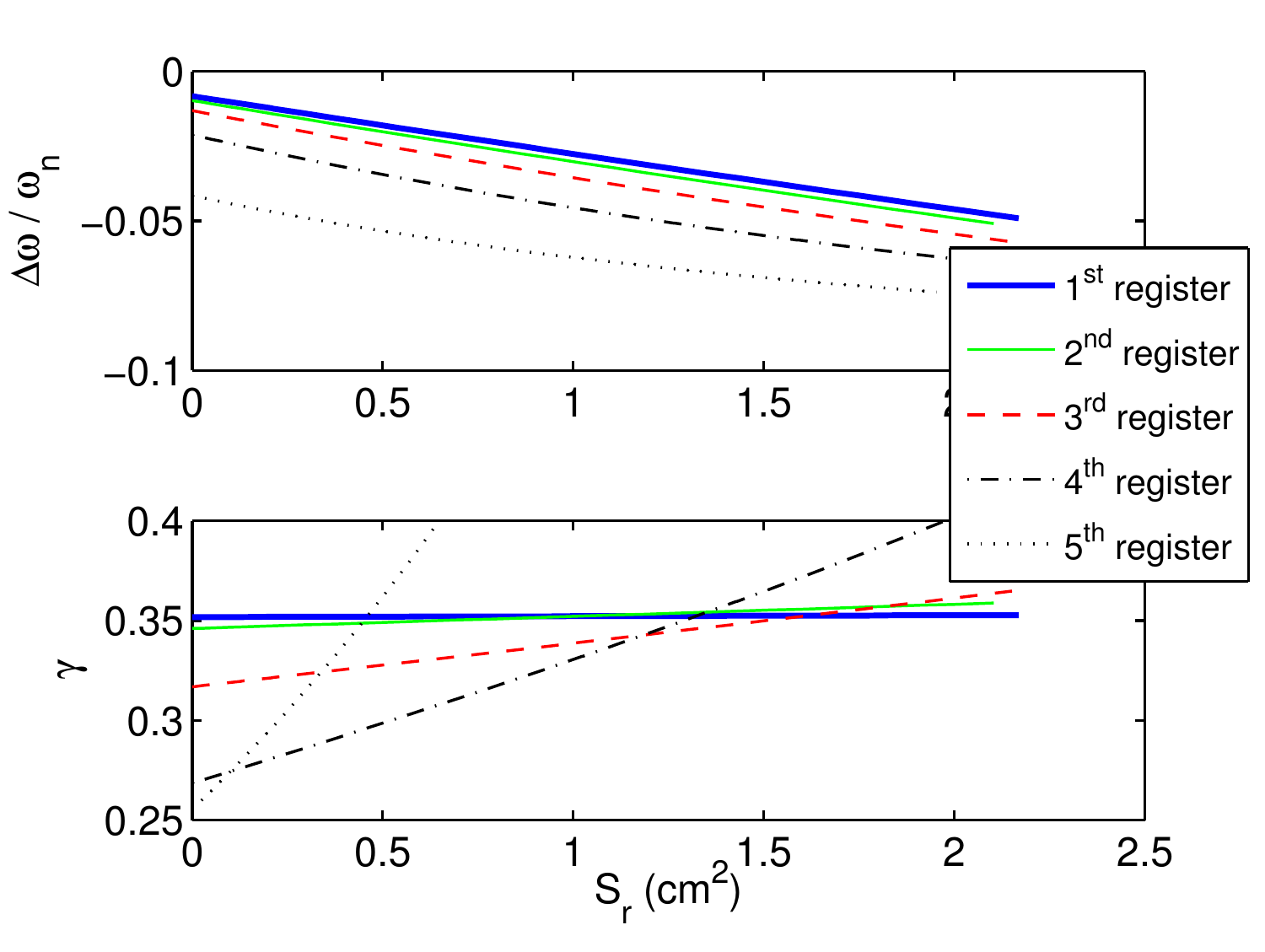}
\caption{\small Variations of the oscillation threshold with respect to $S_r$: critical blowing pressure $\g$ and relative frequency deviation from the bore resonances $\Delta \w/\w_n$ for each of the five Hopf bifurcations. Parameters of the model: $q_r=0.3$, $\zeta=0.75$, $f_r=1500Hz$.\label{fig:2D_ThetaGamma_versus_Sa}}
\end{figure}
Figure \ref{fig:2D_ThetaGamma_versus_Sa} shows the influence of the flow induced by the motion of the reed, through the value of the equivalent area $S_r$ of the reed that participate to this flow. Previous works by Dalmont\cite{dalmont:1995}, Kergomard\cite{kergomard:livre}, or Silva\cite{silva:2008,silva:2009} report values around $\Delta L$=$10mm$, which corresponds, for a reed stiffness $K_a$=$8.10^6Pa/m$ and a bore of radius $r$=$7mm$, to $S_r$=$0.86$cm\tss{2}. Thus we scaled the variations of $S_r$ between $0$ and $2$cm\tss{2}, using parameters values corresponding to a relaxed embouchure, in order to well illustrate its influence.

An important frequency deviation is shown as $S_r$ increases, as well as a regime selection ``cascade'': the selected register is successively the fifth, the fourth, the third, and finally for high values of $S_r$ the first. Finally, comparing with figure \ref{fig:3D_gamma_zeta_qr} (in the plane $\zeta$=$.75$), the reed induced flow acts in a similar manner as the reed damping on the blowing pressure threshold and the regime selection.

\section{Conclusion\label{sec:ccl}}
In this paper, a clarinet physical model is investigated using numerical continuation tools. To the authors' knowledge, it is the first time that such method is used to compute the oscillation threshold variations with respect to several model parameters. The reed dynamics and its induced flow are shown to have critical influence on the regime selection and the minimal blowing pressure necessary to bifurcate from the static regime and establish a steady-state periodic oscillation: a note. Previous works already gave useful insights concerning the influence of some model parameters on ease of play and intonation. The present work confirm and extend these results to the case of a more complex model. Moreover, the method used allows to investigate variations of two parameters at the same time, instead of one, like in previous studies.

In this approach, the parameters of the model are assumed to be constant or to undergo quasi-static variations. However, in real situation, the player can modify some control parameters (e.g. blowing pressure, lip stress on the reed) at a timescale that might sometimes be comparable to the oscillations period. Guillemain\cite{guillemain:2010} reported measured variations of $\g$ on a time scale of a few milliseconds only, which is comparable to the period of an oscillation at $150Hz$. However, despite such a limitation, the results provided through numerical continuation are out of reach for direct time simulations.

Whereas the main results presented here concern a clarinet model, it should be noted that the method itself is very general. No hypotheses (other than linear behaviour) is made on the resonator. Thus, other resonators can be studied by fitting the modal decomposition to its input impedance spectrum. For instance, extending to the case of the saxophone or taking into account the tone holes only requires to compute the corresponding modal decomposition of the input impedance. Even the physical model can be modified. It only has to be written as a set of first order ordinary differential equations (and additional algebraic equations, if necessary). The method could also be used to compare models with each other.

In comparison with the method of the characteristic equation, the computations are much faster: Silva reported ten minutes of computation per branch, whereas it lasts only a few seconds in the present case. Moreover, the continuation algorithm used is very robust: strong variations (as variations of $\g$ for low $k_rL$ in fig. \ref{fig:suivi_krL.eps}) do not require special care and are computed in a straightforward way.

The same continuation method applied to the continuation of periodic solutions, as described by the authors in a conference paper\cite{karkar:ica2010}, also allows to compute the entire dynamic range of a given model of wind instrument, without any additional simplification, unlike previous works (see Dalmont\cite{dalmont:2000} et al). From that perspective, the numerical continuation approach seems promising for the global investigation of the behaviour of a given physical model of musical instrument.

Applications of this work to instrument making are possible: for instance, modifications of the geometry of the resonator can be studied in terms of their influence on ease of play and intonation. Other applications concerning mapping strategies for sound synthesis are also of interest. Indeed, the estimation of the parameters of a model is a difficult task, especially when the parameters values should vary to reproduce through sound synthesis typical behaviours of the instrument modelled. On the one hand, direct measurements on a real player are most of the time highly complicated (for example, $\omega_r$, $q_r$ or even $h_0$ when the lips of the player is pressed on the reed). On the other hand, inverse problems still appear to be limited to the estimation of parameter values from a synthesised sound.

Thus, the approach presented in this paper offers the possibility to know in advance the influence the parameter values on some key features of the model behaviour around the oscillation threshold : ease of play, regime selection and fine intonation. Therefore, mapping strategies could be developed for sound synthesis applications. They would consist in  binding different parameters in order to let a player modify one of them while maintaining the same playing frequency (at threshold) for a given note for  example, or while preserving the ease of play on a whole register.

\bibliography{oscillation-threshold}
\end{document}